\documentstyle[pra,aps,preprint]{revtex}
\begin{document}

\draft
\title{Quantum Hall effect at half-filled Landau level: Pairing of composite fermions}
\author{Takao Morinari}
\address{Department of Applied Physics, University of Tokyo,
Bunkyo-ku, Tokyo 113, Japan}

\date{\today}
\maketitle
\begin{abstract}
We discuss the possibility of the quantum Hall effect 
at half-filled Landau level in terms of the pairing of the 
composite fermions.
In the absence of Coulomb energy, we show that 
the ground state of the system is
described by the {\it p}-wave BCS pairing state of composite fermions.
When the ratio $\alpha \equiv (e^2/\epsilon \ell_B)/\epsilon_F$
($\ell_B$ : the magnetic length, $\epsilon_F$ : Fermi energy of the composite
fermions) is larger than a critical value $\alpha_c \sim 8.2$
the gap of the pairing state vanishes.
However, $\alpha$ remains less than $\alpha_c$ if
$\hbar \omega_c \gg e^2/\epsilon \ell_B$ holds.
Then in this situation it is possible that the pairing state which 
results in the quantum Hall effect occurs.
The effect of the real spin degrees of freedom and the Zeeman energy is
also discussed.
\end{abstract}

The two-dimensional electron systems with partially filled Landau level 
have rich structures.
One of them is the fractional quantum Hall effect
observed at {\it odd} integer denominator 
Landau level filling factor \cite{fqh}.
The ground state of this system 
is the {\it incompressible} liquid \cite{laughlin}. 
Contrary to these {\it odd} denominator filling fraction, 
the quantization of the Hall conductance is 
{\it not} observed at {\it even} integer 
denominator filling fraction \cite{exp,saw} except for the case of 
$\nu=5/2$ \cite{5over2,haldane_rezayi}. 
The possibility of the quantum Hall effect
at these filling fraction is still controversial problem \cite{baskaran}.

One of the theoretical framework to understand the system with $\nu=1/2$
is the composite fermion (CF) picture \cite{jain}.
In this theory, electrons are mapped into CFs 
which have charge $e$ and the fluxes 
$2 \Phi_0$ ($\Phi_0=ch/e$).
Halperin, Lee, and Read (HLR) \cite{hlr}
studied this CF problem within the random phase approximation.
At the mean field level, the fictitious fluxes attached to CFs cancel 
the external magnetic field and the system is described as the fermion system
in the absence of the magnetic field.
The system is, however, not an ordinary Fermi liquid 
due to the fluctuation of the Chern-Simons gauge field.
Including this gauge field, the effective mass of CFs
seems to diverge \cite{hlr}. 
On the other hand, Greiter, Wen, and Wilczek (GWW) discussed the possibility of
the pairing state of CFs at $\nu=1/2$ \cite{greiter}.
However, GWW used an approximation that the number of CF's fluxes is small 
and they retained first order term for Chern-Simons gauge filed and neglected
the quadratic term and the Coulomb interaction term.
Hence how neglected terms affect the pairing state is unclear and
the condition of quantum Hall effect is not discussed \cite{bonesteel}.

In this paper, we discuss the possibility of the pairing state of CFs
taking into account {\it all} interaction terms in the absence of the disorder.
First, we derive the Hamiltonian which takes into account {\it all} interaction
terms for CFs. 
From this Hamiltonian, we derive the gap equation for the spinless CFs 
and analyzing it, 
we show that the ground state of the system is the {\it p}-wave BCS \cite{bcs}
pairing state of CFs.
With increasing the ratio $\alpha \equiv (e^2/\epsilon \ell_B)/\epsilon_F$,
where $\ell_B$ is the magnetic length and $\epsilon_F$ is the Fermi energy
of the composite fermions,
the gap of the pairing state goes to zero.
The pairing state occurs in the region of $\alpha < \alpha_c$,
where $\alpha_c \sim 8.2$.
The value of $\alpha$ is calculated by estimating the Fermi energy 
$\epsilon_F$ contained as the parameter of CF theory.
Using the result of HLR, $\alpha$ is $\sim 6.7$
if the condition $\hbar \omega_c \gg e^2/\epsilon \ell_B$ holds.
This value of $\alpha$ is less than $\alpha_c$. Hence if the condition
$\hbar \omega_c \gg e^2/\epsilon \ell_B$ holds, the pairing state of CFs 
which results in the quantum Hall effect occurs. 
In GaAs samples, it will be realized
in more strong magnetic field than one in present experiments.
We also discuss the effect of the real spin degrees of freedom 
and the Zeeman energy and there we naturally understand the polarization of 
real spin.

To begin with, we derive the effective Hamiltonian for the BCS pairing of CFs.
The second quantized Hamiltonian 
for the two-dimensional spinless electron system
in the presence of an external uniform magnetic field perpendicular to 
the layer is given by
\begin{equation}
H  = \int d^2 {\bf r} \psi^{\dagger} ({\bf r}) 
\left[ \frac{1}{2m} (-i \nabla + {\bf A})^2 - \mu \right]
           \psi ({\bf r}) + V_C, 
\end{equation}
where,
$V_C  =  1/2 \int d^2 {\bf r}_1 \int d^2 {\bf r}_2 
        \left( 1/\epsilon |{\bf r}_1-{\bf r}_2| \right)
        \delta \rho ({\bf r}_1)
        \delta \rho ({\bf r}_2)$.
Here we take the units $\hbar=c=e=1$ and 
$\delta \rho({\bf r})= \psi^{\dagger}({\bf r}) 
\psi({\bf r}) - {\bar \rho}$ 
with ${\bar \rho}$ the average particle density.
We introduce the generalized CF field operators \cite{rajaraman} by
\begin{equation}
\phi ({\bf r}) = e^{-J ({\bf r})} \psi ({\bf r}),
\hspace{10mm} \pi ({\bf r}) = \psi^{\dagger} ({\bf r}) 
                         e^{J ({\bf r})},
\label{non_unitary1}
\end{equation}
where
\begin{equation}
J ({\bf r}) = 2 \cdot \int d^2 {\bf r}^{\prime} 
                        \rho ({\bf r}^{\prime}) \log (z-z^{\prime})
                        -\frac{1}{4{\ell_B}^2} |z|^2,
\label{non_unitary2}
\end{equation}
with $z=x+iy$.
The factor 2 in the first term of r.h.s. of Eq. (\ref{non_unitary2})
denotes the number of fluxes attached to CFs.
The operators 
$\phi ({\bf r})$ and $\pi ({\bf r})$
satisfy the fermion anticommutation relations.

In terms of these operators, the Hamiltonian is described by
\begin{equation}
H = \int d^2 {\bf r} \pi ({\bf r}) \left(
\frac{1}{2m} \left[ (-i \nabla)^2 + \left\{ -i \nabla, \delta {\bf a}
({\bf r}) \right\}
+ \left\{ -i \nabla, i \hat{z} \times \delta {\bf a} ({\bf r})
\right\} \right] -\mu
\right) \phi ({\bf r}) + V_C,
\label{cf_hamiltonian}
\end{equation}
where 
\begin{equation}
\delta {\bf a} ({\bf r}) = 2 \int d^2 {\bf r}^{\prime} 
\left( \pi ({\bf r}^{\prime}) \phi ({\bf r}^{\prime}) - \bar{\rho} \right)
\nabla {\rm Im} \log (z-z^{\prime}),
\end{equation}
and $\{A,B\}=AB+BA$.
Note that we have an irregular term 
which contains $\{-i \nabla, i \hat{z} \times \delta {\bf a}\}$,
because we perform a {\it non-unitary} transformation described by
Eq. (\ref{non_unitary1}) and Eq. (\ref{non_unitary2}).
After performing the Fourier transformations, we obtain 
the effective Hamiltonian for the BCS-pairing of CFs;
\begin{equation}
H_{CF} = \sum_{{\bf k}} \xi_k \pi_k \phi_k 
+ \frac{1}{2 \Omega} \sum_{{\bf k}_1 \neq {\bf k}_2} V_{{\bf k}_1 {\bf k}_2}
\pi_{{\bf k}_1} \pi_{-{\bf k}_1} \phi_{- {\bf k}_2} \phi_{{\bf k}_2},
\end{equation}
where $\xi_k=k^2/2m - \mu$ and,
\begin{equation}
V_{{\bf k}_1 {\bf k}_2} = - \frac{8 \pi i}{m} \cdot 
\frac{{\bf k}_1 \times {\bf k}_2}{|{\bf k}_1 - {\bf k}_2|^2}
- \frac{4 \pi}{m} \cdot \frac{k_1^2 - k_2^2}{|{\bf k}_1 - {\bf k}_2|^2}
+ \frac{\pi \alpha}{4m} \cdot \frac{k_F}{|{\bf k}_1 - {\bf k}_2|},
\label{interaction}
\end{equation}
with $\alpha = (e^2/\epsilon \ell_B)/\epsilon_F$.
The second term of the r.h.s. of Eq. (\ref{interaction}) is derived from the 
irregular term mentioned above.
As we will show later, this term breaks the particle-hole symmetry.

Now we discuss the pairing of CFs. Though we derive the gap equation by the
mean field theory, we replace the electron band mass with the CF's effective
mass $M$ in order to take into account the renormalization effect.
The gap equation of the ground state is given by
\begin{equation}
\Delta_{\bf k} = - \frac{1}{2 \Omega} \sum_{{\bf k}^{\prime} (\neq {\bf k})}
V_{{\bf k} {\bf k}^{\prime}} 
\frac{\Delta_{ {\bf k}^{\prime} } }{E_{k^{\prime}}},
\label{gap_equation}
\end{equation}
and
\begin{equation}
\overline{\Delta}_{\bf k} = 
- \frac{1}{2 \Omega} \sum_{{\bf k}^{\prime} (\neq {\bf k})}
{V_{{\bf k} {\bf k}^{\prime}}}^*
\frac{\overline{\Delta}_{ {\bf k}^{\prime} } }{E_{k^{\prime}}},
\label{gap_equation2}
\end{equation}
where $E_k = \sqrt{\xi_k + \overline{\Delta}_{\bf k} \Delta_{\bf k}}$.

To analyze the $\ell$-wave paring state,
we set 
$\Delta_{\bf k} = |\Delta_k| \exp \left({-i \ell \theta_{\bf k}}\right)$,
where $\ell$ is an integer and $\theta_{\bf k}$ denotes the angle 
of vector ${\bf k}$.
Because we discuss the pairing state of spinless fermions, $\ell$ must be
taken as an odd integer.
Substituting this expression into Eq. (\ref{gap_equation}), we obtain
\begin{equation}
|\Delta_k| = - \frac{1}{2M} 
\left( \int_0^k dk^{\prime} + \int_k^{\infty} dk^{\prime} \right)
\frac{k^{\prime} |\Delta_{k^{\prime}}|}{E_{k^{\prime}}} 
I_{\ell} (k,k^{\prime}),
\end{equation}
where $I_{\ell} (k,k^{\prime})$
is given by
\begin{eqnarray}
I_{\ell} (k,k^{\prime}) 
& = & \int_0^{2 \pi} \frac{d \theta}{2 \pi i}
\left[ 2 \cdot 
\frac{e^{-i \ell \theta} \sin \theta}
{(k^2+{k^\prime}^2)/2k k^{\prime} - \cos \theta}
- i \cdot \frac{k^2-{k^{\prime}}^2}{k k^{\prime}} \cdot 
\frac{e^{-i \ell \theta}}{(k^2+{k^\prime}^2)/2k k^{\prime} - \cos \theta}
\right. \nonumber \\
&   & \left. \hspace{10mm} + \,
\frac{i}{8} \alpha k_F \cdot \frac{e^{-i \ell \theta}}
{\sqrt{k^2 + {k^{\prime}}^2 - 2 k k^{\prime} \cos \theta}} \right].
\end{eqnarray}
To get the nonzero $|\Delta_k|$, $I_{\ell}(k,k^{\prime})$ must be positive.
This is possible only in the case of $\ell > 0$, where
\begin{equation}
I_{\ell} (k,k^{\prime}) 
= \left\{ \begin{array}{cc}
-4 \left( \frac{k}{k^{\prime}} \right)^{\ell} 
+ \frac{\alpha}{8} \cdot \frac{k_F}{k} \cdot
\frac{k^{\prime}/k}{\sqrt{1-(k^{\prime}/k)^2}}
\cdot \frac{1}{\sqrt{1-(k^{\prime}/k)^2}+1} 
\hspace{5mm} & for \hspace{5mm} k>k^{\prime}, \\
\frac{\alpha}{8} \cdot \frac{k_F}{k} \cdot
\frac{k/k^{\prime}}{\sqrt{1-(k/k^{\prime})^2}}
\cdot \frac{1}{\sqrt{1-(k/k^{\prime})^2}+1}
\hspace{5mm} & for \hspace{5mm} k<k^{\prime}.
\end{array} \right.
\end{equation}
The difference of the behavior of $I_{\ell}(k,k^{\prime})$ between
the regime of $k<k^{\prime}$ and $k>k^{\prime}$ 
is caused by the second term
of the r.h.s. of Eq. (\ref{interaction}).
The reason why we encounter this {\it non-symmetric} 
behavior of the gap equation is 
that we perform the {\it non-unitary} transformation.

To solve the gap equation, we introduce an approximation for $|\Delta_k|$;
\begin{equation}
|\Delta_k| = \left\{
\begin{array}{cc}
\Delta_{k_F} \left( k/k_F \right)^{\ell} & \hspace{3mm} 
for \hspace{3mm} k < k_F, \\
0 & \hspace{3mm} for \hspace{3mm} k>k_F. \end{array} \right.
\label{approx_gap}
\end{equation}
With this approximation, the gap equation is transformed into 
\begin{equation}
2 \int_0^1 dx \frac{x^{\ell}}{\sqrt{(x-1)^2 + \Delta^2 x^{\ell}}}
- \frac{\alpha}{8} \int_0^1 dx 
\frac{(1-x^2)^{\ell}}{\sqrt{x^4 + \Delta^2 (1-x^2)^{\ell}}}
\cdot \frac{1}{x+1} = 1,
\label{l_wave}
\end{equation}
where $\Delta \equiv \Delta_{k_F}/\epsilon_F$.

On the other hand,
the ground state energy difference ($\equiv \delta E$)
between the pairing state and no-pairing state
is given by
\begin{equation}
\delta E = \frac{1}{2} \sum_{\bf k} |\xi_k| 
\left( 1-\frac{|\xi_k|}{E_k} \right) 
- \frac{1}{4} \sum_{\bf k} \frac{|\Delta_k|^2}{E_k}.
\label{gs_energy}
\end{equation}
We can show numerically that $\delta E \leq 0$ for $\Delta \geq 0$
and the larger $\Delta$ gives the lower $\delta E$.

Fig. \ref{a_gap} shows the $\alpha$ dependence of the gap $\Delta$.
In the region of $\alpha < \alpha_c$, where $\alpha_c \sim 8.2$, 
the pairing state which has the largest
$\Delta$ is the component of $\ell = 1$.
Hence the ground state of the system
is the {\it p}-wave BCS pairing state in this region 
because the largest gap state has the lowest energy as mentioned above.
On the other hand, in the region of 
$\alpha > \alpha_c$ all of the gaps vanish.
Hence there are no pairing state.
The answer to the question which case must be applied to the real system 
depends on the value of $\epsilon_F$.
In the Sec. IV of Ref. \cite{hlr}, HLR 
estimated the effective mass using a dimensional analysis,
which holds in the limit of $\hbar \omega_c \gg e^2/\epsilon \ell_B$
and numerically obtained gaps of several fractional quantum Hall states.
Using their analysis, $\alpha$ is estimated to $\sim 6.7$.
This value of $\alpha$ is lower than $\alpha_c$.
The ratio of $e^2/\epsilon \ell_B$ to $\hbar \omega_c$ is given by
$(e^2/\epsilon \ell_B)/\hbar \omega_c \sim 4.85 \times 10^2 \times 
(m_b/m_e)/(\epsilon \sqrt{B})$ where $m_b$ is the band mass and 
the external magnetic field is measured in the unit of tesla. 
Hence the pairing state of CFs is possible in the sample with 
small band mass $m_b$, large dielectric constant $\epsilon$ and under the
strong magnetic field $B$ where $\hbar \omega_c \gg e^2/\epsilon \ell_B$ 
holds.

Here we mention the importance of the second term of the r.h.s. of Eq. 
(\ref{interaction}), which is absent in the analysis of GWW. 
If we neglect this term and use an approximation
Eq. (\ref{approx_gap}) but $|\Delta_k| = \Delta_{k_F} (k_F/k)^{\ell}$
for $k>k_F$, we obtain $\alpha_c \sim 5.8$, which is lower than the
value $\alpha \sim 6.7$ obtained in the case of $\hbar \omega_c \gg 
e^2/\epsilon \ell_B$. That is the qunatum Hall effect never occurs.
Hence the second term of the r.h.s. of Eq. 
(\ref{interaction}) must be retained for the proof of the quantum Hall effect
at $\nu=1/2$.

Next we discuss the effect of the real spin degrees of freedom
and the Zeeman energy. First we discuss the former 
in the absence of the latter. The spin unpolarized pairing state is possible
as in the case of the bilayer quantum Hall systems \cite{morinari}. 
In this pairing state, the expression of the ground state energy is 
$\delta E$ times 2 because of the spin degrees of freedom. 
However, Eq.(\ref{gs_energy}) contains the Fermi wave number 
as the parameter in the explicit and implicit way 
and $\delta E$ is proportional to $k_F^4$.
In the spin unpolarized pairing case, the Fermi wave number
$k_F$ is equal to $k_F^p/\sqrt{2}$, where $k_F^p$($=1/\ell_B$) is the 
Fermi wave number of the spin polarized case.
Putting it all together, the ground state energy of the spin unpolarized 
pairing state is half of $\delta E$ estimated in the case of spin polarized
case.
 Being $\delta E <0$, the spin 
polarized state is preferred over the spin unpolarized pairing state.
As a result, it is enough to consider the pairing of spinless CFs
\cite{bilayer}.
With regard to the effect of the Zeeman energy to the spinless CFs,
it is nothing but shifting the chemical potential.

In summary, using the non-unitary transformation we derived the Hamiltonian 
which takes into account {\it all} interaction terms for CFs.
The gap equation and the expression for the ground state energy were
derived within the weak coupling theory.
The pairing state occurs if the condition $\alpha \equiv (e^2/\epsilon \ell_B)/
\epsilon_F < \alpha_c (\sim 8.2)$ is satisfied. This situation is realized
in the sample with small band mass $m_b$, 
large dielectric constant $\epsilon$ and under the strong magnetic field $B$ 
where $\hbar \omega_c \gg e^2/\epsilon \ell_B$ holds.
The symmetry of this pairing state is 
the {\it p}-wave and real spin is polarized.

I would like to thank N. Nagaosa, H. Aoki and M. Ogata for useful discussions.
This work is supported by Research Fellowships of the Japan Society for the 
Promotion of Science for Young Scientists.

\begin{figure}
\caption{The $\alpha(\equiv (e^2/\epsilon \ell_B)/\epsilon_F)$ dependence of
the gap $\Delta$ for the $\ell$-wave pairing state. 
In the region of $\alpha < \alpha_c$, where $\alpha_c \sim 8.2$,
the $\ell$ with the largest gap is $\ell=1$. Hence the {\it p}-wave pairing
of CFs occurs. On the other hand, in the region of $\alpha > \alpha_c$,
there are no pairing state.}
\label{a_gap}
\end{figure}


\begin{references}
\bibitem{fqh} D. C. Tsui, H. L. Stormer, and A. C. Gossard, Phys. Rev. Lett.
{\bf 48}, 1559 (1982).
\bibitem{laughlin} R. B. Laughlin, Phys. Rev. Lett. {\bf 50}, 1395 (1983).
\bibitem{exp} R. Willett, J. P. Eisenstein, H. L. St\"{o}rmer, D. C. Tsui,
A. C. Gossard, and J. H. English, Phys. Rev. Lett. {\bf 59}, 1776 (1987);
H. W. Jiang, H. L. Stormer, D. C. Tsui, L. N. Pfeiffer, and K. W. West,
Phys. Rev B, {\bf 40}, 12013 (1990).
\bibitem{saw} R. Willett, M. A. Paalanen, R. R. Ruel, K. W. West, 
L. N. Pfeiffer, and D. J. Bishop, Phys. Rev. Lett. {\bf 65}, 112 (1990).
\bibitem{5over2} R. Willett, J. P. Eisenstein, H. L. Stormaer, D. C. Tsui,
A. C. Gossard, and J. H. English, Phys. Rev. Lett. {\bf 59}, 1776 (1987).
\bibitem{haldane_rezayi} F. D. M. Haldane and E. H. Rezayi, 
Phys. Rev. Lett. {\bf 60}, 956 (1988).
\bibitem{baskaran} G. Baskaran, Physica B {\bf 212}, 320 (1995).
\bibitem{jain} J. K. Jain, Phys. Rev. Lett. {\bf 63}, 199 (1989);
Phys. Rev. B {\bf 40}, 8079 (1989); {\it ibid}. {\bf 41}, 7653 (1990).
\bibitem{hlr} B. I. Halperin, P. A. Lee, and N. Read, Phys. Rev. B {\bf 47},
7312 (1993).
\bibitem{greiter} M. Greiter, X. G. Wen and F. Wilczek, Nucl. Phys. B 
{\bf 374}, 567 (1992).
\bibitem{bonesteel} See also, N. E. Bonesteel, Phys. Rev. B {\bf 48}, 
11484 (1993).
\bibitem{bcs} J. Bardeen, L. N. Cooper, and J. R. Schrieffer, Phys. Rev. 
{\bf 108}, 1175 (1957).
\bibitem{rajaraman} R. Rajaraman, and S. L. Sondhi, Int. J. Mod. Phys. B
{\bf 10}, 793 (1996); R. Rajaraman, Phys. Rev. B {\bf 56}, 6788 (1997).
\bibitem{morinari} Takao Morinari, unpublished.
\bibitem{bilayer} The discussion here is also applicable to the bilayer quantum
Hall systems at $\nu=1/2$. In that case, the system approaches the
pseudo spin polarized system with decreasing the layer separation $d$ and
we do not observe the quantum Hall effect at $d=0$ as long as 
$\hbar \omega_c \sim e^2/\epsilon \ell_B$.
\end{references}
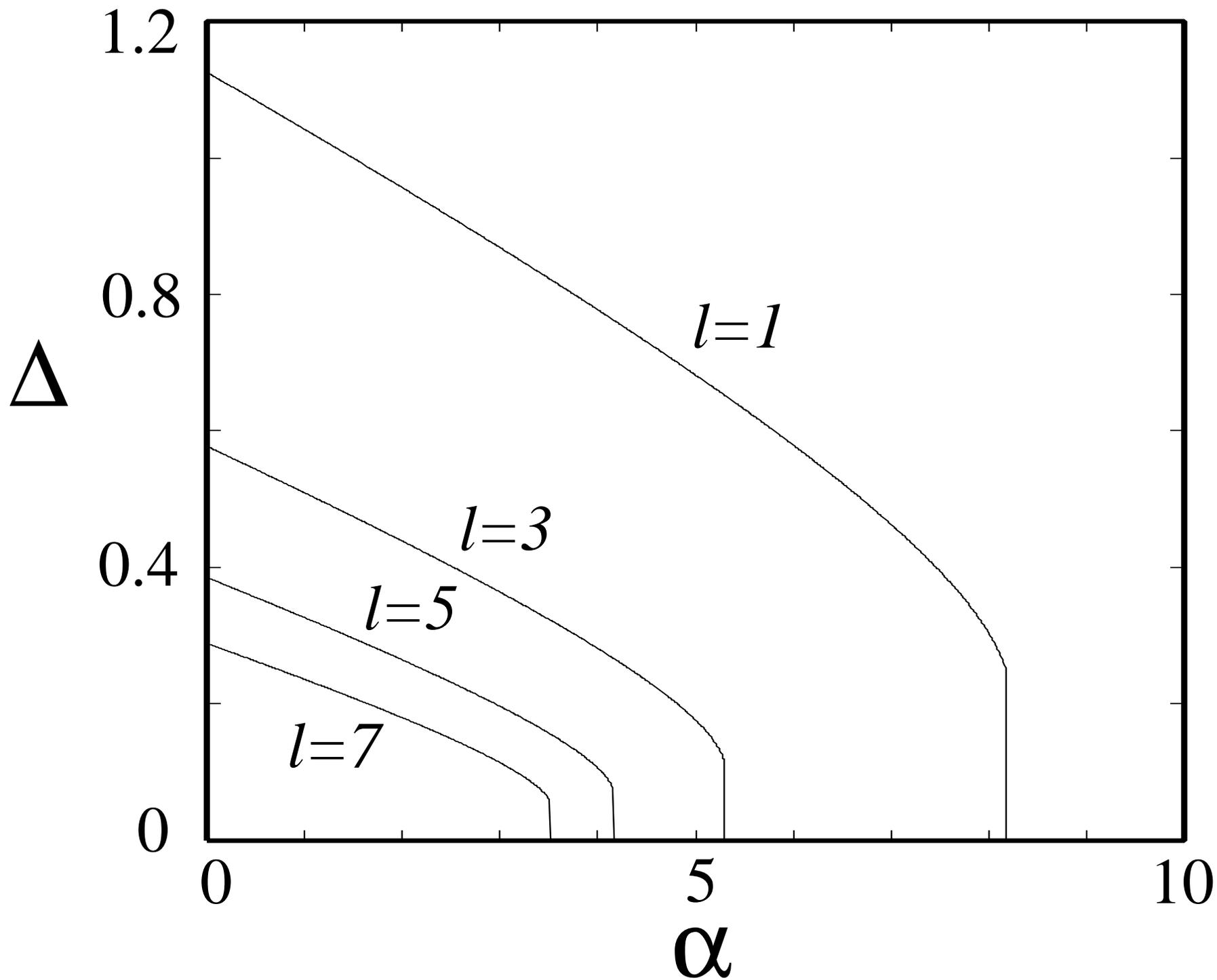
\end{document}